\documentclass[amsmath,amssymb,twocolumn,showpacs,superscriptaddress,prb]{revtex4-1}

\usepackage{dcolumn}% Align table columns on decimal point
\usepackage{bm}% bold math
\usepackage{graphicx}
\usepackage{epstopdf}
\usepackage{wrapfig}
\usepackage{hyperref}
\usepackage{array} 
\usepackage{listings}
\usepackage[para,online,flushleft]{threeparttablex}
\usepackage{booktabs,dcolumn}

\usepackage{tikz}
\usetikzlibrary{arrows}
\usetikzlibrary{shadings}

\usepackage{textpos}
\usepackage{booktabs}
\usepackage{multirow,bigdelim}
\usepackage{float}

\usepackage{xcolor}
\definecolor{pastelgray}{rgb}{0.81, 0.81, 0.77}
\definecolor{beaublue}{rgb}{0.9, 0.9, 0.93}

\begin{document}

\preprint{COLLAPS/Ca spin and moments}

%\author{{R.F. Garcia Ruiz \thanks{KU Leuven}}
\begin{titlepage}
% \hspace{-1cm}
{\fontsize{26}{10} \textbf{\textcolor{black}{\flushleft Unexpectedly large charge radii\\of neutron-rich calcium isotopes}}}\\
{
R.F. Garcia Ruiz$^1$, 
M. L. Bissell$^{1,2}$, 
K. Blaum$^{3}$, 
A. Ekstr\"{o}m$^{4,5}$, 
N. Fr\"{o}mmgen$^{6}$, 
G. Hagen$^{4}$, 
M. Hammen$^{6}$, 
K. Hebeler$^{7,8}$, 
J.D. Holt$^{11}$, 
G.R. Jansen$^{4,5}$, 
M. Kowalska$^{9}$, 
K. Kreim$^{3}$, 
W. Nazarewicz$^{4,10,12}$, 
R. Neugart$^{3,6}$, 
G. Neyens$^{1}$, 
W. N\"{o}rtersh\"{a}user$^{6,7}$, 
T. Papenbrock$^{4,5}$, 
J. Papuga$^{1}$, 
A. Schwenk$^{7,8}$, 
J. Simonis$^{7,8}$, 
K.A. Wendt$^{4,5}$ and 
D.T. Yordanov$^{3,13}$}

{
\fontsize{6}{10}{
\selectfont
$^{1}$KU Leuven, Instituut voor Kern-en Stralingsfysica, B-3001 Leuven, Belgium 
$^{2}$School of Physics and Astronomy, The University of Manchester, Manchester M13 9PL, United Kingdom
$^{3}$Max-Planck-Institut f\"{u}r Kernphysik, D-69117 Heidelberg, Germany 
$^{4}$Physics Division, Oak Ridge National Laboratory, Oak Ridge, Tennessee 37831, USA 
$^{5}$Department of Physics and Astronomy, University of Tennessee, Knoxville, Tennessee 37996, USA 
$^{6}$Institut f\"{u}r Kernchemie, Universit\"{a}t Mainz, D-55128 Mainz, Germany 
$^{7}$Institut f\"{u}r Kernphysik, Technische Universit\"{a}t Darmstadt, 64289 Darmstadt, Germany 
$^{8}$ExtreMe Matter Institute EMMI, GSI Helmholtzzentrum f\"{u}r Schwerionenforschung GmbH, 64291 Darmstadt, Germany 
$^{9}$CERN, European Organization for Nuclear Research, Physics Department, CH-1211 Geneva 23, Switzerland 
$^{10}$Department of Physics and Astronomy and NSCL/FRIB Laboratory, Michigan State University, East Lansing, Michigan 48824, USA 
$^{11}$TRIUMF, 4004 Wesbrook Mall, Vancouver, British Columbia, V6T 2A3, Canada 
$^{12}$Institute of Theoretical Physics, Faculty of Physics, University of Warsaw, Pasteura 5, PL-02-093 Warsaw, Poland
$^{13}$Institut de Physique Nucl\'eaire d'Orsay, CNRS/IN2P3, Universit\'e Paris-Sud, F-91406 Orsay Cedex, France.
}
}
\vspace{0.1cm}
\end{titlepage}

\textbf{Despite being a complex many-body system, the atomic nucleus exhibits simple structures for certain ``magic'' numbers of protons and neutrons. The calcium chain in particular is both unique and puzzling: evidence of doubly-magic features are known in $^{40,48}$Ca, and recently suggested in two radioactive isotopes, $^{52,54}$Ca. Although many properties of experimentally known Ca isotopes have been successfully described by nuclear theory, it is still a challenge to predict their charge radii evolution. Here we present the first measurements of the charge radii of $^{49,51,52}$Ca, obtained from laser spectroscopy experiments at ISOLDE, CERN. The experimental results are complemented by state-of-the-art theoretical calculations. The large and unexpected increase of the size of the neutron-rich calcium isotopes beyond $N = 28$ challenges the doubly-magic nature of $^{52}$Ca and opens new intriguing questions on the evolution of nuclear sizes away from stability, which are of importance for our understanding of neutron-rich atomic nuclei.}

Similar to the electrons in the atom, protons and neutrons (nucleons)
in atomic nuclei occupy quantum levels that appear in ``shells''
separated by energy gaps. The number of nucleons that completely fill
each shell define the so called ``magic'' numbers, known in stable
nuclei as $2, 8, 20, 28, 50, 82, 126$ \cite{goeppert49}. A fundamental
understanding of how these magic configurations evolve in unstable
neutron-rich nuclei, and how they impact the charge radius, is one of
the main challenges of modern experimental and theoretical nuclear
physics
\cite{fridmann05,jones10,holt12,hagen12,wienholtz13,nature54Ca,kim13,ekstrom2015}.

The calcium isotopic chain ($Z=20$) is a unique nuclear system to
study how protons and neutrons interact inside the atomic nucleus: two
of its stable isotopes are magic in both their proton and neutron
number ($^{40}$Ca and $^{48}$Ca), while experimental evidence of
doubly-magic features in two short-lived calcium isotopes have been
reported recently, based on precision measurements of nuclear masses
for $^{52}$Ca ($N=32$) \cite{wienholtz13} and $2^{+}$ excitation
energies for $^{54}$Ca ($N=34$) \cite{nature54Ca}. Possibly,
additional doubly-magic isotopes might exist even further away from
stability \cite{gade14}. As a local change in the behaviour of the charge radius is expected in doubly-magic nuclei
\cite{angeli09}, it is important to pin down the
charge radius in these exotic isotopes to understand how shell
structure evolves and impacts the limits of stability.

\begin{figure*}[t]
\begin{center}
\includegraphics[scale = 0.65]{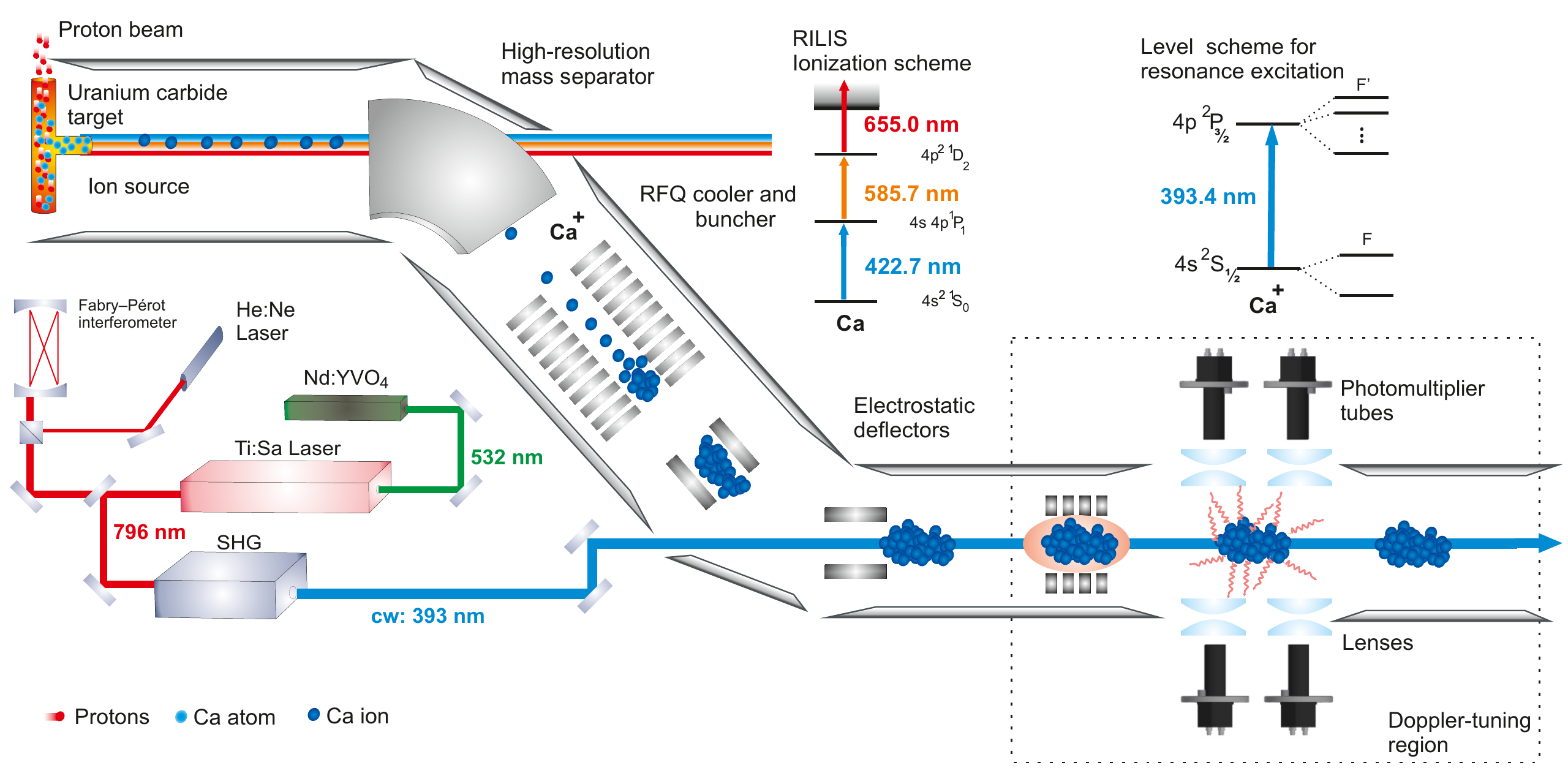}
\caption{
\label{setupISOLDE}
High-resolution bunched collinear laser spectroscopy at ISOLDE, CERN.
Short-lived calcium isotopes are produced from nuclear reactions of
high-energy protons impacting on an uranium carbide target. Ca atoms
were selectively ionized by using a three-step laser scheme
\cite{marsh14}. Ions were extracted from the source and mass separated
to be injected into a radiofrequency trap, where they are accumulated for typically 100 ms. Bunches of ions were
extracted and redirected into the COLLAPS beam line to perform
collinear laser spectroscopy experiments. At COLLAPS, the ions are
superimposed with a continuous wavelength laser beam to scan the
hyperfine structure in the $4s$ $^2S_{1/2}$ $\rightarrow$ $4p$
$^2P_{3/2}$ transition of Ca$^+$ (see text for more details).}
\end{center}
\end{figure*}

\begin{figure}[b]
\includegraphics[scale=0.38]{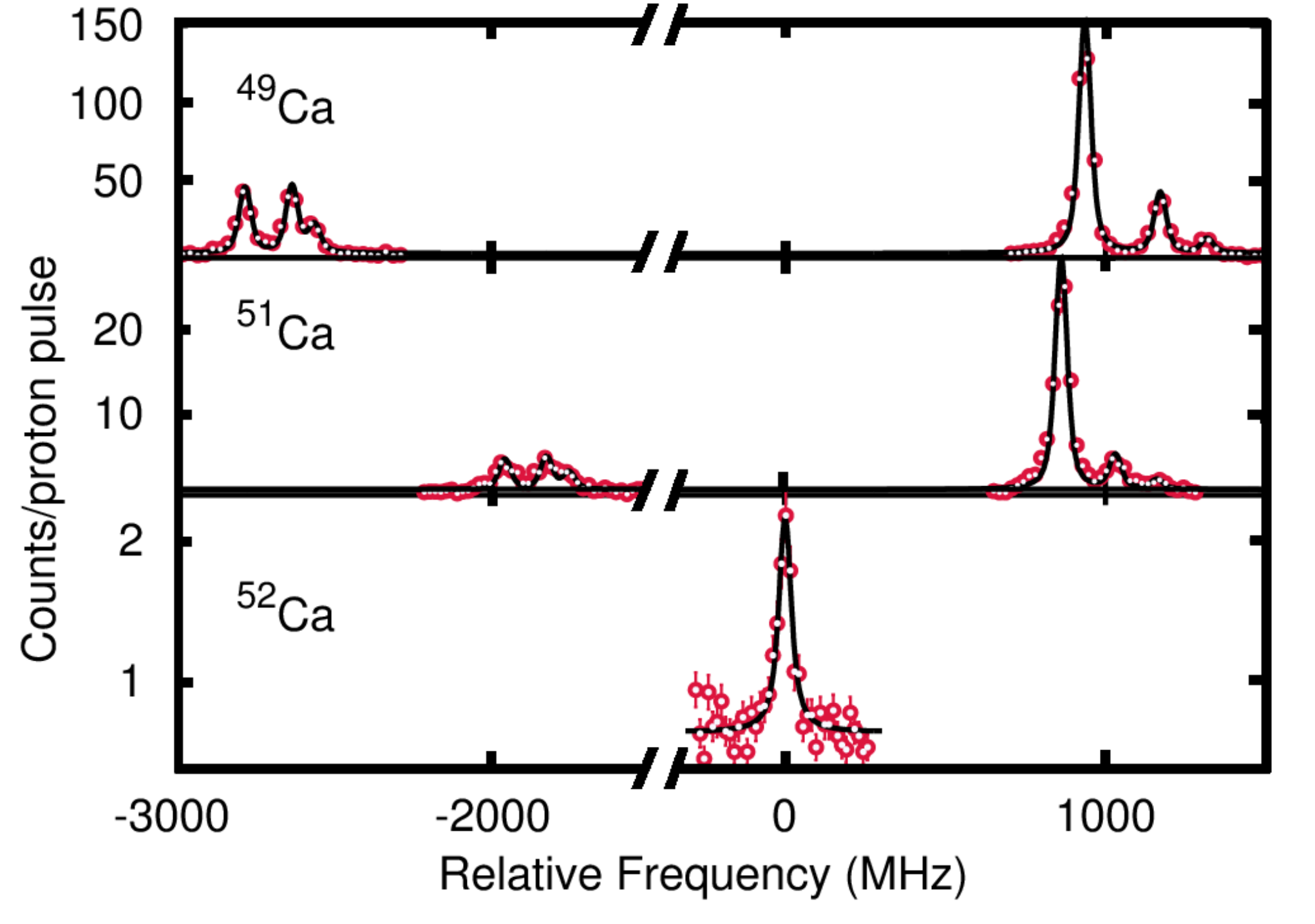}
\vspace{-0.44cm}
\caption{Examples of hyperfine structure spectra measured for the Ca
  isotopes in the 393-nm $4s$ $^2S_{1/2} \rightarrow 4p$ $^2P_{3/2}$
  ionic transition. The solid lines show the fit with a Voigt
  profile. Frequency values are relative to the centroid of
  $^{52}$Ca.\label{spectra}}
\end{figure}

Although the average distance between the electrons and the nucleus in
an atom is about 5000 times larger than the nuclear radius, the size
of the nuclear charge distribution is manifested as a perturbation of
the atomic energy levels. A change in the nuclear size between two
isotopes gives rise to a shift of the atomic hyperfine structure
levels.  This shift between two isotopes, commonly known as
the isotope shift, $\delta \nu^{A,A'}$, includes a part that is
proportional to the change in the nuclear mean square charge radii,
$\delta \langle r^2 \rangle^{A,A'}$ (see expression (\ref{eq:radii})
in methods). Isotope shifts of stable Ca isotopes have been extensively
studied in the literature \cite{palmer}, revealing the unusual
evolution of their charge radii. Despite an excess of eight neutrons,
$^{48}$Ca exhibits the striking feature that it has essentially the
same charge radius as $^{40}$Ca. A fundamental explanation of these
anomalous features has been a long-standing problem for nuclear theory
for more than three decades (see, e.g.,
\cite{caurier01,bender03}). Quantitative scenarios have been proposed
to account for some of these properties \cite{talmi84,fayans00,caurier01}, but
so far a microscopic description was lacking.

A first estimate for the charge radius of the radioactive $^{49}$Ca
isotope was based on a failure to observe its resonance in a
dedicated measurement of its isotope shift. The resulting conclusion that $\delta \nu^{48,49} \leq 50$\,MHz suggested an enormous
increase of the charge radius $\delta \langle r^2 \rangle ^{48,49}
\geq 0.5$ fm$^2$ \cite{andl82}, reflecting the strong magicity of
$^{48}$Ca. So far, the only charge radius measured beyond $^{48}$Ca
has been for $^{50}$Ca, resulting in a large increase of
$\delta \langle r^{2}\rangle^{48,50} =0.293(37)$ fm$^2$
\cite{vermeeren}. These results raised even more exciting questions on
the charge radii evolution of Ca isotopes. It suggested that the
prominent odd-even staggering of their charge radii could be even more pronounced beyond
$^{48}$Ca since a reduction of the charge radius for $^{52}$Ca would
be expected as a consequence of a suggested doubly-magic nature of
this isotope.  Thus, the experimental determination of the charge
radii of $^{49,51,52}$Ca not only addresses fundamental questions
regarding the size of atomic nuclei, but are also important for
understanding the possible doubly-magic character of $^{52}$Ca. By
using high-resolution bunched-beam collinear laser spectroscopy at
ISOLDE, CERN, changes in the charge radii for $^{40-52}$Ca isotopes
were obtained from the measured optical isotope shifts (see Table
\ref{chargeradii_S}). With a production yield of only a few hundred
ions per second, our measurement of $^{52}$Ca represents one of the
highest sensitivities ever reached by using fluorescence detection
techniques.

\begin{table*}[t]
\begin{center}
%\begin{minipage}{\textwidth}
%\begin{center}
\caption{\textbf{Measured isotope shifts and corresponding rms charge radii.}
Statistical and systematic uncertainties are given in round and
squared brackets, respectively.}
%\begin{tabular}{{l}@{\extracolsep{1cm}}lcccccccc}
\begin{tabular}{@{\extracolsep{0.6cm}}l*{2}{D{.}{.}{7}}cl*{4}{D{.}{.}{7}}@{}}
\hline
\multicolumn{3}{c}{$\delta \nu^{40,A}$ (MHz)} & \multicolumn{3}{r}{$\delta \langle r^2\rangle$ (fm$^2$)}\\
\cline{2-4} 
\cline{5-7}
%\multicolumn{2}{c}{This work} & \multicolumn{5}{c}{Literature}\\
\multicolumn{1}{r}{A} &\multicolumn{1}{l}{This work} & \multicolumn{2}{l}{Literature}&\multicolumn{1}{l}{This work} & \multicolumn{1}{r}{Literature}\\
%\multicolumn{2}{c}{This work} & \multicolumn{3}{c}{Literature}\\
\cline{3-4} 
\cline{5-5}
\cline{6-7}
%\cmidrule(r){2-3} 
%\cmidrule{2-3}
% & This work & & Literature\\ 
%$A$ && &$\delta \nu^{40,A}$ (MHz) &$\delta \langle r^2\rangle$ (fm$^2$)    & $\delta \langle r^2\rangle$ (fm$^2$)&$\delta \langle r^2\rangle$ (fm$^2$)  \\
\cline{2-2} 
\cline{3-3}
\cline{4-4} 
\cline{5-5}
\cline{6-6}
40 & 0 & 0\footnote{High-precision measurements of isotope shifts of stable Ca isotopes \cite{triga15}.} & 0\footnote{Data from Ref \cite{vermeeren}.} &0\footnote{Values of $K_{SMS}=-8.8(5)$ GHz~u and $F=-276(8)$ MHz/fm$^2$ were taken from the King-plot results (see methods). In square brackets are listed the systematic errors associated to these factors, which are related to the uncertainty of the kinetic energy of the ion beam.} & 0\footnote{Recalculated from isotope shift values reported in \cite{vermeeren} using the newly reported mass values \cite{AME}.} & 0\footnote{Taken from Ref. \cite{palmer}.}\\
42&-&426.4(15)[10]&  425(4)[4]	     &   &	0.203(15) &0.215(5)	\\
43&683.0(12)[16]& &672(9)[6]  &0.114(4)[8] &0.125(30) &0.125(3)	\\
44&851.1(6)[21]&850.1(10)[20]&842(3)[8]&0.288(2)[6] 	&	0.280(11)	& 0.283(6)\\
45&1103.5(7)[25]&&1091(4)[10]&0.125(3)[8]	&	0.126(15) & 0.119(6)	\\
46&1301.0(6)[30]&&1287(3)[12]&0.125(2)[8]	&	0.124(11) & 0.124(5)	\\
47&1524.8(8)[35]&& &0.002(3)[9]	& 	 & 0.005(13)	\\
48&1706.5(8)[38]&1710.6(35)[42]&1696(4)[16]&0.001(3)[10]	&	-0.022(15)& -0.004(6)	\\
49&1854.7(10)[43]&& &0.098(4)[12]	&			\\
50&1969.2(9)[47]&&1951(9)[20]&0.291(3)[12]	&	0.276(34)	\\
51&2102.6(9)[51]&& &0.392(3)[13]	&			\\
52&2219.2(14)[56]&& &0.531(5)[15]	&			\\
 \hline
\end{tabular}
%In square brackets the errors associated to $K_{SMS}$
\label{chargeradii_S}
%\end{center}
%\end{minipage}
\end{center}
\end{table*}

The isotope shift is typically $10^{6}$ times smaller than the
absolute transition frequency between two atomic levels, and the
relevant part is in turn only a small fraction of the total shift. Thus
a measurement of such a tiny change is only possible by using ultra-high
resolution techniques.  Collinear laser spectroscopy has been
established as a unique method to reach such high resolution and has
been applied with different detection schemes to study a variety of
isotopes \cite{geithner00,cheal10,blaum13}.

A major challenge of exotic nuclei is that they are produced
in very low quantities and exist only for a fraction of a
second. Therefore, continuous improvements have been made during the
past decades to enhance the experimental sensitivity, without
sacrificing resolution. Since the introduction of an ion
cooler-buncher for collinear spectroscopy that allows bunching radioactive beams \cite{nieminem02}, the sensitivity of the
conventional fluorescence laser spectroscopy has been improved by two
to three orders of magnitude. It is now possible to routinely perform
experiments with beams of $\sim 10^{4}$ ions/s \cite{vingerhoets10}.
%For the very specific case of $^{50}$Ca, an alternative particle detection technique was used to improve the limits of sensitivity \cite{vermeeren}.
In this work, we have further optimized the photon detection
sensitivity and at the same time reduced further the photon background
events \cite{kim13}, now allowing to study calcium isotopes produced
with a yield of only a few hundred ions per second.
%Thus, our measurement of $^{52}$Ca represents now one of the highest sensitivities ever reached by optical detection techniques. 
While preserving the high resolution, this sensitivity surpasses the
previous limit by two orders of magnitude, achieved by an
ultra-sensitive particle detection technique employed on Ca isotopes
\cite{vermeeren}.
 
The short-lived Ca isotopes studied in this work were produced at the
ISOLDE on-line isotope separator, located at the European Center for
Nuclear Research, CERN. High-energy proton pulses with intensities of $\sim 3 \times 10^{13}$ protons/pulse at 1.4 GeV
impinged every 2.4 seconds on an uranium carbide target to create radioactive species of
a wide range of chemical elements. The Ca isotopes were selected from
the reaction products by using a three-step laser ionization scheme
provided by the Resonance Ionization Laser Ion Source (RILIS)
\cite{marsh14}. A detailed sketch of the different experimental
processes from the ion beam production to the fluorescence
detection is shown in Fig.~\ref{setupISOLDE}.

After selective ionization, Ca ions (Ca$^+$) were extracted from the
ion source and accelerated up to 40 keV. The
isotope of interest was mass-separated by using the High-Resolution Mass Separator (HRS). The selected isotopes were
injected into a gas-filled radiofrequency trap (RFQ) to accumulate the
incoming ions. After a few milliseconds, bunches of ions of $\sim
5\mu$s temporal width were extracted and redirected into a dedicated
beam line for collinear laser spectroscopy experiments (COLLAPS). At
COLLAPS, the ion beam was superimposed with a continuous wave laser beam fixed at 393-nm wavelength (see methods), close to
the $4s\ ^2S_{1/2}$ $\rightarrow$ $4p\ ^2P_{3/2}$ transition in the
Ca$^+$.  The laser frequency was fixed to a constant value, while the
ion velocity was varied inside the optical detection region.  A change in the ion velocity corresponds to a
variation of laser frequency in the ion rest frame. This Doppler
tuning of the laser frequency was used to scan the hyperfine structure
(hfs) components of the $4s\ ^2S_{1/2}$ $\rightarrow$ $4p\ ^2P_{3/2}$
transition. At resonance frequencies, transitions between different hfs levels
were excited, and subsequently the fluorescence photons were detected
by a light collection system consisting of four lenses and photomultipliers tubes
(PMT) (see Refs.~\cite{kim13} for
details). The photon signals were accepted only when the ion bunch
passed in front of the light collection region, reducing the
background counts from scattered laser light and PMT dark counts by a
factor of $\sim$ 10$^{4}$. A sample of the hfs spectra measured during
the experiment is shown in Fig.~\ref{spectra}.

The isotope shifts were extracted from the fit of the hfs experimental
spectra, assuming multiple Voigt profiles in the $\chi^2$-minimization
(see methods). The measured isotope shift relative to the reference
isotope $^{40}$Ca, and the corresponding change in the mean-square
charge radius are shown in Table \ref{chargeradii_S}. Statistical
errors (parentheses) correspond to the uncertainty in the
determination of the peak positions in the hfs spectra. The systematic
errors in the isotope shift (square brackets) are mainly due to the
uncertainty in the beam energy, which is also the main contribution to
the uncertainty in the charge radius. Independent high-precision
measurements of isotope shifts on stable Ca isotopes \cite{triga15}
were used for an accurate determination of the kinetic energy of each
isotope.  The stability of the beam energy was controlled by measuring
the stable $^{40}$Ca, before and after the measurement of each isotope
of interest.

\begin{figure}[t]
\begin{center}
\includegraphics[width=0.9\columnwidth]{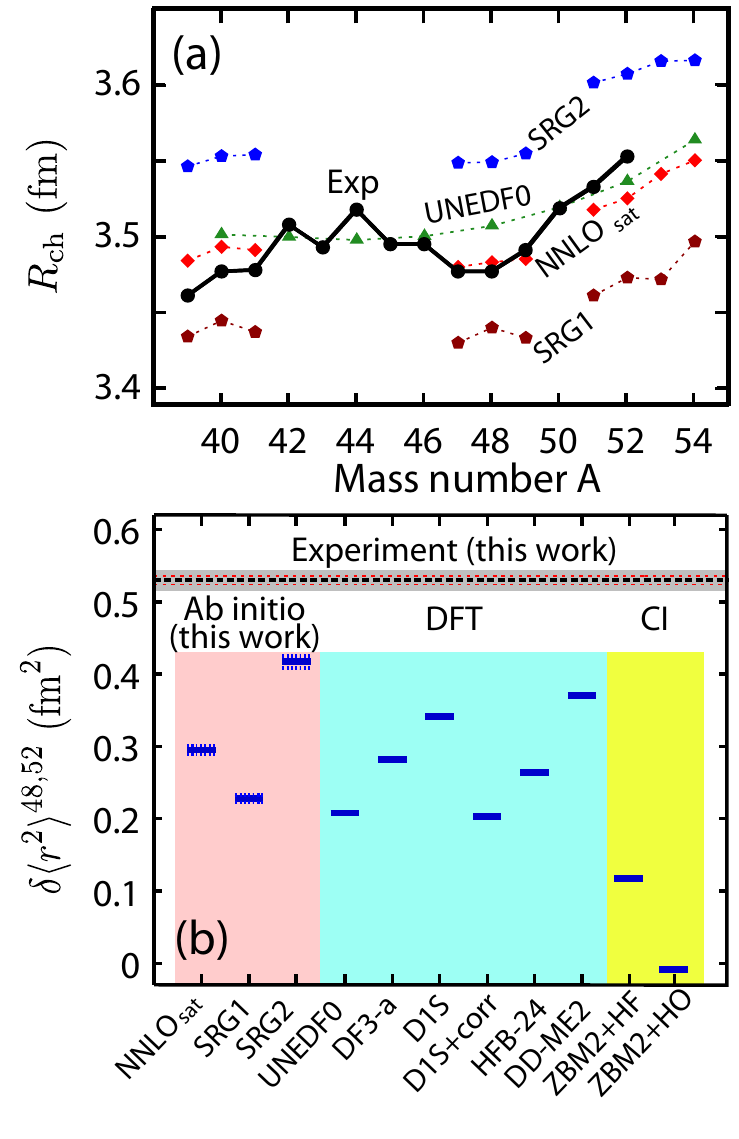}
\end{center}
\vspace{-0.9cm}
\caption{Charge radii of Ca isotopes. (a) Experimental charge radii
  compared to ab initio calculations with chiral EFT interactions
  NNLO$_{\rm sat}$, SRG1, SRG2, as well as DFT calculations with the
  UNEDF0 functional. Experimental error bars are smaller than the
  symbols. The absolute values were obtained from the reference radius
  of $^{40}$Ca ($R_{\rm ch}=3.478(2)$ fm) \cite{fricke}. The values of
  $^{39}$Ca and $^{41,42}$Ca are taken from Ref. \cite{vermeeren96}
  and Ref. \cite{martensson}, respectively. A systematic theoretical
  uncertainty of 1\% is estimated for the absolute values due to the
  truncation level of the coupled-cluster method and the finite
  basis-space employed in the computation. (b) Experimental rms charge
  radius in $^{52}$Ca relative to that in $^{48}$Ca compared to the ab
  initio results as well as those of representative density functional
  theory (DFT) and configuration interaction (CI) calculations. The
  systematic uncertainties in the theoretical predictions are
  largely cancelled when the differences between rms charge radii are
  calculated (dotted horizontal blue lines). Experimental uncertainties are represented by the
  horizontal red lines (statistical) and the gray shaded band
  (systematic).\label{Caradii}}
\end{figure}

Our experimental results (Table \ref{chargeradii_S} and
Fig.~\ref{Caradii}) show that the root-mean-square (rms) charge radius of $^{49}$Ca
presents a considerable increase with respect to $^{48}$Ca, $\delta
\langle r^2\rangle^{48,49}=0.097(4)$ fm$^2$, but much smaller than
previously suggested \cite{andl82}.  The increase continues towards
$N=32$, resulting in a very large charge radius for $^{52}$Ca, with an increase relative to $^{48}$Ca of $\delta \langle
r^2\rangle^{48,52}=0.530(5)$ fm$^2$.
%which is similar to the large increase observed for open-shell nuclei like Fe
%($N=26$) \cite{fricke}, where there is not a sizable shell gap at $N=32$.
%No experimental information is available to compare the charge radii among doubly-magic nuclei in another isotopic chain, but the increase due to the addition of 4 neutrons to $^{48}$Ca is comparable with the relative increase observed between the semi-magic and doubly-magic nuclei, $^{112}$Sn and $^{132}$Sn, which have 20 neutrons difference\cite{blanc05}. 
This increase observed beyond the neutron number $N=28$ is as large as
the values observed for open-shell nuclei like Fe \cite{fricke}, where
there is not a sizable shell gap at $N=32$. Thus, the charge radius of
$^{52}$Ca is found to be much larger than expected for a doubly-magic
nucleus.

With advances in chiral effective field theory (EFT) and the
development of powerful many-body methods, nuclear structure theory
has entered a new era in recent years. Chiral EFT allows to
systematically derive nuclear forces in terms of low-energy degrees of
freedom, nucleons and pions, based on the symmetries of the
fundamental theory, Quantum Chromodynamics. Chiral EFT provides
systematically improvable Hamiltonians, explains naturally the
hierarchy of two-, three-, and higher-body forces, and allows to
estimate theoretical uncertainties~\cite{epelbaum09}.

Ab initio calculations based on chiral EFT interactions are able to
correctly predict properties of light nuclei~\cite{navratil07} and
oxygen isotopes \cite{Herg13oxygen}. In addition, many-body
calculations starting from a $^{40}$Ca core provide an excellent
description of the binding energies of neutron-rich Ca
isotopes~\cite{wienholtz13}, $2^+$ excitation
energies~\cite{holt12,hagen12,coraggio09} and ground-state electromagnetic
moments \cite{garciaruiz15}. However, there are also clear indications
for deficiencies of some chiral EFT interactions: ground-state
energies of heavier nuclei turn out to be systematically too large and
radii are significantly too small~\cite{binder14,ekstrom2015}.

Very recently, this problem was addressed by an optimization of chiral
EFT interactions \cite{ekstrom2015} that also included binding
energies and radii of selected nuclei up to mass number $A = 25$ in
addition to data from two- and few-body systems. Ab initio
coupled-cluster calculations based on the resulting interaction
(NNLO$_{\rm sat}$) accurately reproduce both charge radii and binding
energies of nuclei up to $^{40}$Ca, and the empirical saturation point
of symmetric nuclear matter. In addition to NNLO$_{\rm sat}$, we employ
two chiral interactions~\cite{Hebe10snm}, SRG1 and SRG2, derived from
renormalization group techniques~\cite{Furn13rgrev} (see methods for
details), which are fit only to properties of $A \leq 4$ and predict
nuclear matter saturation within theoretical uncertainties.

In Fig.~\ref{Caradii} we compare the experimental results to the
 charge radii obtained in our ab initio calculations and to other theoretical predictions.
Figure~\ref{Caradii}~(a) shows that our
calculations correctly yield similar charge radii for $^{40}$Ca and
$^{47,48}$Ca for the chiral EFT interactions employed. These
calculations are based on the single-reference coupled-cluster method,
which is ideally suited for nuclei with at most one or two nucleons
outside a closed (sub-) shell~\cite{hagen2014}. Thus, we do not give
theoretical results for the mid-shell isotopes $^{42-46}$Ca and
$^{50}$Ca. We note that absolute values of charge radii are very well
reproduced by NNLO$_{\rm sat}$. The interactions SRG1 and SRG2 also
reproduce well the overall trend, but as they were not optimized to
saturation properties can give either somewhat too low or too high
saturation densities, corresponding to larger or smaller charge
radii. Also shown are nuclear DFT results obtained with the Skyrme
energy density functional UNEDF0 \cite{kortelainen10}, which fails to
describe the fine details of the experimental trend.

Figure~\ref{Caradii}~(b) shows the difference in rms charge radii
between $^{52}$Ca and $^{48}$Ca predicted with different methods and
models; all being representative of modern approaches to charge
radii. In general, for neutron-rich isotopes beyond $^{48}$Ca, our ab
initio calculations consistently predict an increase in charge radii
for $^{50,52}$Ca but fail short of describing the data.  Similarly,
DFT predictions with various models
\cite{erler12,saper11,delaroche2010,rocamaza2011,goriely2013,rossi2015};
configuration interaction (CI) calculations \cite{rossi2015} obtained
from large-scale shell-model calculations with the ZBM2 interaction
\cite{caurier01,bissell14} using Skyrme-Hartree-Fock (ZBM2+HF) and 
harmonic oscillator (ZBM2+HO) wave functions, all considerably
underestimate the large charge radius of $^{52}$Ca. The standard
explanation involving quadrupolar correlations \cite{caurier01,rossi2015} does
not seem to improve this, as can be seen by comparing the D1S and D1S
plus quadrupolar correlations (D1S+corr) results \cite{delaroche2010}
in Fig.~\ref{Caradii}~(b). (For more discussion of the impact of
dynamical quadrupole and octupole effects on charge radii in the Ca chain, see also Ref.~\cite{fayans00}.)
Thus, our experimental results are truly
unexpected. Speculating about the reason for these theoretical
shortcomings we note that all theoretical approaches are lacking in
the description of deformed intruder states associated with complex
configurations.

To assess the impact of core breaking effects, which turned out to be
important for the description of electromagnetic moments in this
region \cite{garciaruiz15}, we studied the proton occupancies of
natural orbitals above the naively filled $Z=20$ shell. Our ab
initio calculations show a weak, but gradual erosion of the proton
core as neutrons are added. While this defies the simple pattern of a rigid proton core expected
for the magic Ca isotopes, the estimated magnitude of core breaking
effects, including coupling to the neutrons, is not sufficient to explain the large charge radius of
$^{52}$Ca.

In summary, the charge radii of $^{49,51,52}$Ca were measured for the
first time, extending the picture for the evolution of charge radii
over three neutron shell closures; a unique scenario found in the entire
nuclear chart. We find that between $^{48}$Ca and $^{52}$Ca the charge
radius exhibits a strong increase that considerably exceeds
theoretical predictions.  This is a surprising finding, considering
that $^{40}$Ca and $^{48}$Ca have similar charge radii
\cite{vermeeren}, which are significantly smaller than the charge
radius of $^{52}$Ca. Our measurements are complemented by ab initio
calculations that reproduce the similarities of charge radii in
$^{40,48}$Ca but cannot account for the unexpected increase in $R_{\rm
ch}$ to $^{52}$Ca. These calculations also show an increase in
proton occupancies outside the $Z=20$ core as more neutrons are added
to $^{40}$Ca, hinting at a progressive weakening of the $Z=20$ shell
closure in the neutron-rich calciums. Future experiments aim at
extending the current studies to isotopes even further away from
stability, especially for the possibly doubly-magic nucleus $^{54}$Ca
\cite{nature54Ca}. These results open intriguing questions on the
evolution of charge radii away from stability and constitute a major
challenge in the search of a unified description of the atomic
nucleus.

\newpage

\noindent
\textbf{\textcolor{blue}{METHODS}}

\textbf{\textcolor{blue}{Laser setup}} The continuous wave laser beam at
393-nm wavelength was obtained from a second-harmonic generation of a Ti:Sa laser output, pumped with 532 nm light from a cw
Nd:YVO$_{4}$ laser. To reduce the laser frequency drift to less than
10 MHz per day, the laser frequency was locked to a Fabry-Perrot
interferometer, which was in turn locked to a polarization-stabilized
HeNe laser (see Figure \ref{setupISOLDE}).

\textbf{\textcolor{blue}{Chiral EFT interactions}} The ab initio
calculations in this work are based on interactions derived within
chiral effective field theory. The interaction NNLO$_{\rm sat}$ includes
contributions to nucleon-nucleon and three-nucleon forces up to third
order in the chiral expansion, which have been fitted to selected
properties of nuclei up to $^{24}$O, so as to include information on
saturation properties (see Ref.~\cite{ekstrom2015} for details). The
interactions SRG1 and SRG2 are derived from the fourth-order chiral
nucleon-nucleon interaction of Ref.~\cite{Ente03EMN3LO} by performing
an evolution to lower resolution scales via the the similarity
renormalization group~\cite{Hebe10snm}. The three-nucleon interactions
include contributions up to third order in the chiral expansion and
are fit to the binding energy of $^3$H and the charge radius of $^4$He
at the low resolution scales. The interaction SRG1 refers to an
interaction at the the resolution scales $\lambda/\Lambda_{\rm 3NF} =
2.8/2.0$ fm$^{-1}$(EM $c_i$'s) and SRG2 to $\lambda/\Lambda_{\rm 3NF} =
2.0/2.0$ fm$^{-1}$(PWA $c_i$'s) (see Table 1 in Ref.~\cite{Hebe10snm}).

\textbf{\textcolor{blue}{Coupled-cluster calculations}} Our
coupled-cluster (CC) calculations are performed starting from a
closed-shell Hartree-Fock reference state computed from $15$
harmonic-oscillator shells with an oscillator frequency of $\hbar \Omega=$22 MeV. In
CC theory correlations are included via a similarity transformation,
which generates particle-hole excitations to all orders in
perturbation theory. In our calculations the cluster operator is
truncated at the singles and doubles level (CCSD), and also includes
triples excitations in a nonperturbative but approximate way (see
Ref.~\cite{hagen2014} for details). Contributions from three-nucleon
forces are taken into account up to $E_{\text{3max}} = 18~\hbar \Omega$ for NNLO$_{\rm sat}$
and $E_{\text{3max}} = 16~\hbar \Omega$ for SRG1 and SRG2 in the normal-ordering
approximation, which has been demonstrated to be an excellent
approximation for medium-mass nuclei~\cite{roth12}. Based on the
size-extensitivity of the CC method, the uncertainty of the radius
results is estimated to 1\% in the CCSD approximation. 
% Begin GH & TP 
The uncertainties of the \emph{ab-initio} results shown in
Fig.~\ref{Caradii}~(b) are very small, because they are differences of
correlated and very small (1\%) uncertainties of each individual
radius. We find 0.006~fm$^2$, 0.005~fm$^2$, and 0.008~fm$^2$ for the
uncertainties of $\delta\langle r^2\rangle^{48,52}$ for the
interactions NNLO$_{\rm sat}$, SRG1, and SRG2, respectively.
% End GH & TP 

The results for the charge radii are computed from the point-proton
radius starting from the intrinsic operator. We then correct the
point-proton radius for the finite proton and neutron charge sizes,
for the relativistic Darwin-Foldy term, and the spin-orbit
correction. The latter is calculated consistently in CC
theory. Finally, we note that the DFT results also include this
spin-orbit correction. If not added in the original reference, we adopt spin-orbit corrections given by
the mean of CC theory and relativistic mean-field theory.

\textbf{\textcolor{blue}{Data analysis}} The measured hyperfine
structure spectra were fitted with Voigt profiles of common widths for
the different hfs components. The hfs centroid obtained for each
isotope were used to extract the isotope shifts, $\delta \nu^{A,A'}$,
and calculate the corresponding change in rms charge radius, $\delta
\langle r^2 \rangle^{A,A'}$, following the expression
\begin{equation}
\small
% \delta \langle r^2 \rangle^{A,A'}=\langle r^2 \rangle^{A'}-\langle r^2 \rangle^{A}=\dfrac{\delta \nu^{A,A'}}{F}-K_{MS}\dfrac{M_{A'}-M_{A}}{F(M_{A'}+m_{e})M_{A}}. 
\delta \langle r^2 \rangle^{A,A'}=\langle r^2 \rangle^{A'}-\langle r^2 \rangle^{A}=\frac{1}{F} \left( \delta\nu^{A,A^\prime} - K_{\rm MS} \frac{M_{A^\prime} - M_A }{M_A M_{A^\prime}} \right).
\label{eq:radii}
\end{equation}
The mass shift, $K_{\rm MS}$, is the sum
of the normal mass shift, $K_{\rm NMS}$, and the specific mass shift,
$K_{\rm SMS}$. The value $K_{\rm NMS}=417.97$~GHz~u 
%%% AS: What is .u in line above?
was obtained from the
energy of the $4s$ $^2S_{1/2}$ $\rightarrow$ $4p$ $^2P_{3/2}$
transition of Ca$^+$. The electronic field factor, $F$, and the
specific mass shift factor, $K_{\rm SMS}$, were obtained by comparing with
independent measurements of charge radii in a King-plot analysis
\cite{palmer}. The results: $F=-276(8)$ MHz/fm$^2$ and
$K_{\rm SMS}=-8.8(5)$ GHz~u, are in good agreement with the literature
values \cite{martensson}.

\textbf{\textcolor{blue}{Acknowledgments}} This work was supported by
the IAP-project P7/12, the FWO-Vlaanderen, GOA grants 10/010 and
15/010 from KU Leuven, the Max-Planck Society, the ERC Grant
No. 307986 STRONGINT, the BMBF contract 05P12RDCIC and 05P15RDCIA, the U.S. Department of Energy, Office of Science, Office of Nuclear
Physics under Award Numbers DEFG02-96ER40963 (University of
Tennessee), DE-SC0013365 (Michigan State University), DE-SC0008499 and
DE-SC0008511 (NUCLEI SciDAC collaboration), the Field Work Proposal
ERKBP57 at Oak Ridge National Laboratory (ORNL), and Contract
No. DE-AC05-00OR22725 (ORNL). Computer time was provided by the
Innovative and Novel Computational Impact on Theory and Experiment
(INCITE) program. This research used resources of the Oak Ridge
Leadership Computing Facility at ORNL, and used computational
resources of the National Center for Computational Sciences, the
National Institute for Computational Sciences. We thank Javier Men\'endez for very useful discussions. We would like to thank
the ISOLDE technical group for their support and assistance.

\textbf{\textcolor{blue}{Author contributions}} R.F.G.R., M.L.B.,
N.F., M.H., M.K., K.K., R.N., W.N\"{o}., J.P., and D.T.Y. performed
the experiment.  R.F.G.R. performed the data analysis and prepared the
figures. G.H. and G.R.J. performed the coupled-cluster calculations. R.F.G.R,
K.B., G.H., K.H., G.N., W.Na., W.N\"{o}., T.P., and A.S. prepared the
manuscript. All authors discussed the results and contributed to the
manuscript at all stages.

\textbf{\textcolor{blue}{Author information}} 
Reprints and permissions information is available at www.nature.com/reprints.
The authors declare no competing financial interests. Readers are welcome to
comment on the online version of the paper. Correspondence and requests for
materials should be addressed to R.F.G.R (ronald.fernando.garcia.ruiz@cern.ch)

\bibliographystyle{naturemag}
\bibliography{bibliography}

\end{document}